\documentstyle[aps,prl,graphicx]{revtex}

\begin{document}
\draft

\twocolumn[\hsize\textwidth\columnwidth\hsize\csname@twocolumnfalse%
\endcsname

\title{The effects of violating detailed balance on critical dynamics}

\author{Uwe C. T\"auber {}$^1$, Vamsi K. Akkineni {}$^{1,2}$, and 
        Jaime E. Santos {}$^{3}$}

\address{$^1$ Department of Physics, Virginia Tech, Blacksburg, VA 24061-0435\\
        $^2$ Department of Physics, University of Illinois at Urbana-Champaign,
        Urbana, IL 61801-3080 \\
       $^3$ Hahn--Meitner-Institut, Abteilung Theorie, D-14109 Berlin, Germany}

%\date{\today}

\maketitle

\begin{abstract}
We present an overview of the effects of detailed-balance violating 
perturbations on the universal static and dynamic scaling behavior near a 
critical point.
It is demonstrated that the standard critical dynamics universality classes are
generally quite robust:
In systems with {\em non-conserved} order parameter, detailed balance is 
effectively restored at criticality.
This also holds for models with {\em conserved} order parameter, and 
{\em isotropic} non-equilibrium perturbations.
Genuinely novel features are found only for models with conserved order 
parameter and spatially {\em anisotropic} noise correlations.
\end{abstract}

]
% 05.40.+j -- Fluctuation phenomena, random processes, and Brownian motion
% 64.60.Ak -- Renormalization-group, fractal, and percolation studies of phase
%             transitions
% 64.60.Ht -- Dynamic critical phenomena

One of the major goals in theoretical non-equilibrium physics has been the
identification and classification of universality classes for the 
long-wavelength, long-time scaling behavior both near continuous dynamical
phase transitions, and for systems displaying generic scale invariance.
Indeed, through investigations of certain specific models, a number of 
prototypical non-equilibrium universality classes have been identified.
Prominent examples are driven diffusive systems \cite{bearev}, models of driven
interfaces and growing surfaces \cite{intrev}, depinning transitions 
\cite{deprev}, and phase transitions from active to absorbing states 
\cite{dirper}, e.g., in diffusion-limited chemical reactions.

A complementary approach is to study the influence of non-equilibrium
perturbations on the known universality classes for equilibrium dynamical
critical phenomena \cite{hohhal}.
Equilibrium critical dynamics is concerned with the relaxational and reversible
kinetics near a thermodynamic critical point at temperature $T_c$, as 
generically described by the Landau-Ginzburg-Wilson (LGW) model for an 
$n$-component order parameter vector field ${\bf S}$ in $d$ space dimensions 
\cite{stafth}.
In addition to the two independent static critical exponents, e.g. the 
correlation length exponent $\nu$ defined via $\xi \propto |\tau|^{-\nu}$ 
($\tau = T-T_c$) and Fisher's exponent $\eta$ for the algebraic decay of the 
two-point correlation function at criticality ($T=T_c$), 
$C(x-x') \propto |x-x'|^{-(d-2+\eta)}$, the order parameter relaxation is 
governed by a dynamic exponent $z$ that describes critical slowing down: 
The characteristic time scale diverges as $t_{\rm ch} \propto |\tau|^{-z \nu}$
upon approaching the transition.
This allows for time scale separation and thus a formulation of critical
dynamics in terms of non-linear Langevin equations:
The relevant `slow' modes consist of the order parameter and all conserved 
quantities to which it is statically or dynamically coupled.
All remaining `fast' degrees of freedom are captured through an effective 
Gaussian white noise.
Different values for $z$ ensue depending on whether the order parameter is a
conserved quantity or not, and on the additional conserved quantities present.
The diffusive relaxation of the latter near criticality can either be 
characterized by the same exponent $z$ as for the order parameter (`strong'
dynamic scaling), or be given by different power laws (`weak' scaling) 
\cite{hohhal}.

In order to ensure relaxation towards thermal equilibrium at long times, as
given by a Gibbs distribution, one has to carefully implement detailed-balance 
conditions.
In the language of non-linear Langevin equations, these are (i) the Einstein
relation between the relaxation constants and the noise strengths, and (ii) the
condition that the probability current associated with reversible kinetics be
divergence-free \cite{intcon}.
Naturally, the following question arises: 
{\em What happens if the equilibrium conditions are violated?}
In this letter, we shall explore two generic types of detailed-balance 
violations, namely (a) coupling the order parameter and additional conserved
quantities to heat baths with different temperatures, and (b) allowing for
spatially anisotropic noise correlations for conserved variables.
To determine their universal features, we map the Langevin dynamics for most of
the models listed in Ref.~\cite{hohhal} to a dynamic field theory 
\cite{janded}, and employ standard renormalization group (RG) methods 
\cite{stafth}.

In the theory of static critical phenomena, an analogous issue concerns the
effect of terms that break the original order parameter symmetry, e.g., the 
influence of cubic anisotropies on the isotropic $n$-component Heisenberg 
model.
There, the rotational symmetry is restored at criticality, provided 
$n < n_c \approx 4$ \cite{cubani}.
Since detailed balance originates from time-reversal symmetry, we might 
anticipate that it could, under certain conditions, effectively become 
reinstated in non-equilibrium critical dynamics, whereupon the asymptotic 
scaling laws are those of the corresponding equilibrium model.
Yet even then, the question arises whether there exist any non-equilibrium 
dynamical RG fixed points that could strongly influence crossover regimes.
The other possible scenario is of course that violating detailed balance 
constitutes a relevant perturbation, rendering the equilibrium RG fixed point 
unstable, and driving the system towards a genuine non-equilibrium stationary 
state.

The simplest case represents {\em purely relaxational dynamics}, with either 
non-conserved or conserved, and therefore diffusively relaxing, order parameter
${\bf S}$ \cite{modsab} ({\em models A and B} according to the classification 
in Ref.~\cite{hohhal}, respectively).
With the effective LGW Hamiltonian
\begin{equation}
  H[{\bf S}] = \int \! d^dx \! \left[ \frac{\tau}{2} \, {\bf S}(x)^2 
  + \frac{1}{2} \left[ \nabla {\bf S}(x) \right]^2
  + \frac{u}{4!} \, {\bf S}(x)^4 \right] \ ,
\label{ham}
\end{equation}
the corresponding Langevin equations read
\begin{equation}
  \partial_t {\bf S}(x,t) = - \lambda \, (i \nabla)^a \, 
  \delta H[{\bf S}] / \delta {\bf S}(x,t) + {\mathbf \zeta}(x,t)
\label{mab}
\end{equation}
where $a=0 \, (2)$ for models A (B).
The noise has zero mean, $\langle {\mathbf \zeta}(x,t) \rangle = 0$, and its
correlations are taken to be
\begin{equation}
  \langle \zeta^\alpha(x,t) \, \zeta^\beta(x',t') \rangle = 2 {\tilde \lambda} 
  \, (i \nabla)^a \, \delta^{\alpha \beta} \delta(x - x') \delta(t - t') \ .
\label{noi}
\end{equation}
The equilibrium Einstein relation requires 
${\tilde \lambda} = k_{\rm B} T \, \lambda$, which ensures that the
associated probability distribution for a configuration ${\bf S}$ approaches 
$P_{\rm eq}[{\bf S}] \propto \exp (- H[{\bf S}] / k_{\rm B} T)$ as 
$t \to \infty$.
Setting ${\tilde \lambda} \not= \lambda$ violates detailed balance; yet this 
simply represents a temperature shift here.
Indeed, upon rescaling ${\bf S} \to ({\tilde \lambda}/\lambda)^{1/2} {\bf S}$ 
and $u \to {\tilde u} = ({\tilde \lambda}/\lambda)^{1/2} u$, detailed balance
is formally restored, albeit with a different non-linear static coupling
\cite{uwezol}.
At criticality, the RG flow will take the latter to the Heisenberg fixed point,
independent of its initial value, and thus the equilibrium critical exponents 
are recovered, namely to lowest non-trivial orders in the deviation
$\epsilon = 4-d$ from the upper critical dimension $d_c=4$: 
$\nu^{-1} = 2 - (n+2) \epsilon / (n+8)$, 
$\eta = (n+2) \epsilon^2 / 2(n+8)^2$ \cite{stafth}, and $z = 2 + c \, \eta$ 
with $c = 6 \ln \frac{4}{3} - 1 + O(\epsilon)$ for model A, whereas 
$z = 4 - \eta$ (exactly) for model B \cite{modsab}.
We already note that the same rescaling with 
$g^2 \to {\tilde g}^2 = {\tilde \lambda} g^2 / \lambda$ maps the 
isotropic $O(3)$-symmetric non-equilibrium model J \cite{uwezol}
\begin{eqnarray}
  &&\partial_t {\bf S} = - g \, \delta H[{\bf S}] / \delta{\bf S} \times 
  {\bf S} + \lambda \nabla^2 \, \delta H[{\bf S}] / \delta{\bf S} + 
  {\mathbf \zeta} \ ,
\label{mjd} \\
  &&\langle \zeta^\alpha(x,t) \zeta^\beta(x',t') \rangle = - 2 
  {\tilde \lambda} \nabla^2 \delta^{\alpha \beta} \delta(x - x') \delta(t - t')
\label{mjn}
\end{eqnarray}
onto its equilibrium version, with $z = (d+2-\eta)/2$ \cite{modelj}.

It is indeed a well-established fact that model-A or Glauber dynamics for the 
kinetic Ising model ($n=1$) is quite robust against non-equilibrium 
perturbations \cite{fhaake,grins1}, even when these break the up/down symmetry 
\cite{kevbea}.
Novel features only arise when Kawasaki dynamics is introduced, whereupon the
order parameter becomes conserved (model B), and in addition the noise strength
is rendered {\em anisotropic}: 
${\tilde \lambda} \nabla^2 \to {\tilde \lambda}_\parallel \nabla_\parallel^2 + 
{\tilde \lambda}_\perp \nabla_\perp^2$ in Eq.~(\ref{noi}), corresponding to
coupling the longitudinal and transverse sectors (of dimensions 
$d_{\parallel / \perp}$) to heat baths with different temperatures 
$T_{\parallel / \perp}$.
For this {\em two-temperature} or {\em randomly driven model B}, and for 
$T_\parallel > T_\perp$, only the transverse sector softens at $T_c$.
As a consequence, the system develops marked anisotropies akin to driven 
lattice gases (or equilibrium uniaxial dipolar magnets, Lifshitz points, and 
ferroelastic materials), described by the generalized scaling law 
$C(q_\parallel,q_\perp,\omega,\tau) = q_\perp^{- 2 - z + \eta} \, {\widetilde 
C}_\pm( q_\perp \xi , q_\parallel / q_\perp^{1 + \Delta} , \omega / q_\perp^z)$
\cite{bearev}. As $\Delta \not= 0$, equilibrium isotropic scaling is violated.
However, upon omitting irrelevant terms, the Langevin equation at criticality 
may be written in the form
\begin{equation}
  \partial_t {\bf S}(x,t) = \lambda_\perp \nabla_\perp^2 \,  
  \delta H_{\rm eff}[{\bf S}] / \delta {\bf S}(x,t) + {\mathbf \zeta}(x,t) \ ,
\label{ttb}
\end{equation}
with an effective Hamiltonian that contains {\em long-range} interactions
\cite{bearoy} [here, $\int_q \ldots = \int \! d^dq / (2 \pi)^d \ldots$]
\begin{equation}
   H_{\rm eff} = \! \int_q \! 
   \frac{c q_\parallel^2 + q_\perp^2 (\tau_\perp + q_\perp^2)}{2 \, q_\perp^2} 
   \, |{\bf S}(q)|^2 + \! \int \!\! d^dx \, \frac{{\tilde u}_\perp}{4!} \, 
   {\bf S}^2 .  
\label{efh}
\end{equation}
These strongly affect the nature of the ordered phase, and reduce the upper 
(to $d_c = 4-d_\parallel$) as well as lower critical dimension 
\cite{bearoy,kevzol}.
To one-loop order (only), the $\epsilon$ expansion for the static critical 
exponents formally yields the same results as for the equilibrium Heisenberg 
model, albeit with $\epsilon = 4-d-d_\parallel$.
The two-loop results for $d_\parallel=1$ can be found in Ref.~\cite{beate2}.
Moreover, the exact scaling relations $\Delta = 1 - \eta/2$ and $z = 4-\eta$ 
hold.

We have extended these considerations to models C and D \cite{uwevam}, which 
take the static coupling 
\begin{equation}
  \Delta H[{\bf S},\rho] = \int \! d^dx \left[ \frac{1}{2} \, \rho(x)^2 
  + \frac{g}{2} \, \rho(x) \, {\bf S}(x)^2 \right] 
\label{cdh}
\end{equation}
of the non-conserved/conserved order parameter to the conserved energy density 
$\rho$ into account \cite{modscd}.
The dynamics is then defined through Eq.~(\ref{mab}) with 
$H[{\bf S},\rho] = H[{\bf S}] + \Delta H[{\bf S},\rho]$, setting $a=0 \, (2)$ 
for model C (D), and 
\begin{eqnarray}
  &&\partial_t \rho(x,t) = D \nabla^2 \, 
  \delta H[{\bf S},\rho] / \delta \rho(x,t) + \eta(x,t) \ ,
\label{mcd} \\
  &&\langle \eta(x,t) \, \eta(x',t') \rangle = - 2 {\widetilde D} \, \nabla^2 
  \, \delta(x - x') \delta(t - t') \ .
\label{cdn}
\end{eqnarray}
In addition to ${\tilde u}$, there appears now a three-point coupling 
$f = {\tilde \lambda} g^2 / \lambda$, also marginal in $d_c=4$, as well as the 
dimensionless ratio of time scales $w = \lambda / D$, and the parameter 
$\Theta = {\widetilde D} \lambda / D {\tilde \lambda}$, which can be 
interpreted as the temperature ratio $T_\rho / T_S$ of the heat baths coupled 
to the energy density and order parameter, respectively.
For {\em model C}, we find the associated RG one-loop beta function
$\beta_\Theta = - 2 \Theta (1 - \Theta) \, f \, w / (1+w)^2$.
This result already establishes the stability of the {\em equilibrium} fixed 
point $\Theta_{\rm eq}^* = 1$.
In fact, we have not found any other, genuinely non-equilibrium RG fixed points
corresponding to $\Theta = 0$ or $\Theta = \infty$ at all.
This remains true even when we allow for spatially anisotropic noise for the
conserved field $\rho$, as the RG beta function for the additional new variable
$\sigma = \Theta_\parallel / \Theta_\perp$ becomes to one-loop order:
$\beta_\sigma = - \sigma (1-\sigma) \, f \, \Theta_\perp / 3 (1+w)^3$, with the
{\em isotropic} fixed point $\sigma_{\rm is}^* = 1$ being stable.
Thus, there are merely the three (one-loop) {\em equilibrium} scaling regimes 
\cite{modscd}:
(a) Strong-scaling, $n=1$: $w^*=1$, $z_S = z_\rho = 2 + \alpha / \nu$;
(b) weak-scaling, $2 \leq n < 4$: $w^*=\infty$, 
$z_S = 2 (1 + \alpha / n \nu) < z_\rho = 2 + \alpha / \nu$;
(c) `model A', $n \geq 4$: $w^*=\infty$, $z_S = 2 + c \eta$, $z_\rho = 2$; here
$\alpha = 2 - d \nu$ denotes the specific-heat critical exponent.
In contrast to the above results, a {\em linear} coupling of a conserved mode 
to the order parameter induces effective long-range interactions, leading to  
novel genuine non-equilibrium behavior \cite{grins2}.

For {\em model D}, the characteristic time scale for the order parameter is 
always much larger than that of the diffusive field, and therefore $w^*=0$.
By integrating out the scalar fields $\rho$ from the dynamic functional, one
can then readily show that the non-equilibrium parameter $\Theta$ disappears
entirely from the theory in the limit $w \to 0$ \cite{uwevam}, which completely
reduces the model to the equilibrium theory, with $z_S = 4 - \eta$, whereas 
$z_\rho = 2 + \alpha / \nu$ for $\alpha > 0$ and $z_\rho = 2$ for $\alpha < 0$
(`model B').
Upon allowing for spatially anisotropic order parameter noise correlations, we
arrive at a {\em two-temperature model D}.
As for the corresponding pure order parameter relaxation dynamics, the
non-equilibrium perturbations then induce strong anisotropies, whereupon the
effective Hamiltonian (\ref{efh}) enters the model-D equations of motion, with
only transverse Laplacians, and the associated downwards shift of $d_c$.
The model-D scaling relations for the anisotropy and dynamic exponents still
apply, as well as the two different regimes for $\alpha > 0$ and $\alpha < 0$, 
albeit with the static exponents $\nu, \eta$, $\alpha$ of the long-range 
anisotropic theory \cite{uwevam}.

Purely relaxational dynamics provides, however, an insufficient description for
many real systems.
Often, fully {\em reversible}, non-dissipative terms originating in the 
microscopic dynamics need to be taken into account; in isotropic ferromagnets
(model J), for example, there is also the spin precession in the local magnetic
field, see Eq.~(\ref{mjd}).
Such reversible contributions in the Langevin equations for critical dynamics
may also involve {\em mode-couplings} to other conserved and therefore slow
variables:
In planar ferromagnets, the (non-conserved) order parameter is confined to the 
$xy$ plane, say, but the non-vanishing commutators of the spin components yield
a coupling to the diffusive fluctuations of the conserved $z$ component of the 
magnetization ($\langle S_z \rangle = 0$).
In isotropic antiferromagnets, the order parameter is represented by the
three-component staggered magnetization (not conserved), dynamically coupled to
the conserved magnetization.
The corresponding Langevin equations define {\em models E} and {\em G}, 
respectively \cite{modseg}; their generalization to $n$ order parameter 
components is termed the {\em SSS model}, originally introduced in the context 
of structural phase transitions \cite{sssmod,modssh}.
Lastly, a consistent description of the critical dynamics near the liquid-gas 
transition, or equivalently, of the phase separation in binary liquids, 
involves not only the conserved scalar order parameter density $S$ (a linear 
combination of the mass and energy densities), but in addition the independent
and also conserved transverse corresponding current density ${\bf j}_\perp$ 
({\em model H}) \cite{modelh,modssh}.
In the following, we describe the effects of non-equilibrium perturbations on 
these dynamic universality classes with reversible mode-couplings 
\cite{uwezol,uwejai}.

The results of our investigations of isotropic and anisotropic 
{\em non-equilibrium} versions of the {\em SSS model} were already reported in 
Ref.~\cite{uwezol}; for completeness we review our essential findings here.
As in model C, we may choose different effective temperatures for the 
$n$-component order parameter ${\bf S}$ and the $n(n-1)/2$ non-critical,
conserved generators ${\bf M}$ of the rotation group $O(n)$.
The coupled Langevin equations of motion read
\begin{eqnarray}
  &&\partial_t S^\alpha = g \sum_{\beta \not= \alpha} M^{\alpha \beta} S^\beta 
  - \lambda \, \delta H[S] / \delta S^\alpha + \zeta^\alpha \ , 
\label{sss} \\
  &&\langle \zeta^\alpha(x,t) \, \zeta^\beta(x',t') \rangle = 2 {\tilde\lambda}
  \, \delta^{\alpha \beta} \delta(x - x') \delta(t - t') \ ,
\label{ssn} \\
  &&\partial_t M^{\alpha \beta} = - g \left( S^\alpha \nabla^2 S^\beta - 
  S^\beta \nabla^2 S^\alpha \right) + D \nabla^2 M^{\alpha \beta} + 
  \eta^{\alpha \beta} \ , \nonumber \\
  && \label{ssm} \\
  &&\langle \eta^{\alpha \beta}(x,t) \, \eta^{\gamma \delta}(x',t') \rangle = 
  - 2 {\widetilde D} \nabla^2 \delta(x - x') \delta( t - t') \nonumber \\
  &&\qquad\qquad\qquad\qquad\qquad \times \left( \delta^{\alpha \beta} 
  \delta^{\gamma \delta} - \delta^{\alpha \delta} \delta^{\beta \gamma} \right)
  \ .
\label{smn}
\end{eqnarray}
As opposed to the purely relaxational models, there exist genuine 
non-equilibrium fixed points here, corresponding to temperature ratios 
$\Theta = T_M / T_S = 0$ and $\infty$.
In the former case, there is no feedback to the order parameter, and the
critical dynamics is essentially model-A like ($z_S = z_M = 2$), with anomalous
noise correlations $\propto q^{d-2}$ for the generators ${\bf M}$.
For $\Theta_N^* = \infty$, the order parameter dynamics does not affect the
conserved fields, thus $z_M = 2$, but is itself strongly influenced by their
fluctuations ($w = \lambda/D = \infty$).
We find $z_S = d/2$, as in equilibrium, but a modified {\em static} exponent
$\nu^{-1} = 2 - 2(n+2)\epsilon/(n+8)$ \cite{uwezol}.
However, both these fixed points are {\em unstable}, and the asymptotic 
critical properties are governed by the {\em equilibrium strong-scaling} fixed
point with $z_S = z_M = d/2$ \cite{sssmod,modssh}.
Allowing for spatially anisotropic noise for the generators ${\bf M}$ provides 
the new parameter $\sigma = \Theta_\parallel / \Theta_\perp$, but does not
change the overall picture:
The isotropic fixed point $\sigma_{\rm is}^*$ remains stable, whence the
equilibrium critical behavior is eventually recovered \cite{uwezol}.
The anisotropic fixed points are essentially determined by a combination of the
$\Theta = 0$ and $\Theta = \infty$ characteristics in the different spatial 
sectors.

The intriguing question now arises how non-equilibrium perturbations affect the
critical dynamics of a {\em conserved} order parameter field, when reversible 
mode-couplings are present.
We have therefore investigated the {\em non-equilibrium model H}, as defined by
\cite{uwejai}
\begin{eqnarray}
  &&\partial_t S = - g (\nabla S) \cdot {\bf j} + \lambda \nabla^2 \left( 
  \tau - \nabla^2 + \frac{u}{6} \, S^2 \right) S + \zeta \ , 
\label{hsd} \\
  &&\langle \zeta(x,t) \, \zeta(x',t') \rangle = - 2 {\tilde \lambda} \ 
  \nabla^2 \, \delta(x - x') \delta(t - t') \ ,
\label{hsn} \\
  &&\partial_t \, {\bf j}_\perp = {\bf T} \left[ g (\nabla S) \left( \tau - 
  \nabla^2 \right) S + D \nabla^2 {\bf j} + {\mathbf \eta} \right] \ , 
\label{hjd} \\
  &&\langle {\mathbf \eta}(x,t) \cdot {\mathbf \eta}(x',t') \rangle = - 2 d \,
  {\widetilde D} \ \nabla^2 \, \delta(x - x') \delta(t - t') \ ,
\label{hjn}
\end{eqnarray}
where ${\bf T}({\vec q}) = \delta_{ij} - q_i \, q_j / q^2$ denotes the 
transverse projector in momentum space, and we have omitted irrelevant (in the 
RG sense) terms.
As in model D, asymptotically $w = \lambda/D \to 0$; as a consequence, only 
{\em one} additional non-equilibrium fixed point $\Theta_N^* = 0$ is allowed.
In analogy with the corresponding SSS model fixed point, the order parameter
dynamics is model-B like, $z_S = 4$, while $z_j = 2$, with anomalous noise
correlations again $\propto q^{d-2}$ in Fourier space \cite{uwejai}.
Yet from 
$\beta_\Theta = - (1-\Theta) \frac{2}{3} {\tilde f}(1 + 1/16 \, \Theta)$, where
${\tilde f} = {\tilde \lambda} g^2 / \lambda^2 D$, we infer that once more the 
{\em equilibrium} weak-scaling fixed point $\Theta_E^* = 1$ is stable, with 
$z_S = 4 - 18 \epsilon/ 19$ and $z_j = 2 - \epsilon/ 19$ 
(Galilean invariance fixes $z_S + z_j = d + 2$) \cite{modelh,modssh}.
However, the generalization to spatially {\em anisotropic} noise, 
${\tilde \lambda} \, \nabla^2 \to {\tilde \lambda}_\parallel \, 
\nabla_\parallel^2 + {\tilde \lambda}_\perp \, \nabla_\perp^2$ and
${\widetilde D} \, \nabla^2 \to {\widetilde D}_\parallel \, \nabla_\parallel^2 
+ {\widetilde D}_\perp \, \nabla_\perp^2 $ has a drastic effect.
As in the corresponding models B and D, in this {\em two-temperature model H}, 
the characteristic anisotropic DDS singularities emerge. 
In combination with the mode-coupling terms, this prevents the system from 
approaching an effective equilibrium model: 
The divergence-free condition for the reversible probability current cannot be 
satisfied.
In fact, to one-loop order at least we even find a run-away flow for the RG 
couplings, and are thus unable to determine the long-time scaling behavior 
\cite{uwejai}.
Remarkably, the similarly constructed anisotropic, {\em two-temperature model 
J} is plagued by the same pathology \cite{uwezol}, as is a recently studied 
{\em uniformly} rather than randomly {\em driven model J} \cite{dasrao}.
In that instance, computer simulations revealed that the system displays 
spatio-temporal chaos at long times; perhaps the absence of an RG fixed point 
in the two-temperature models J and H may indicate chaotic behavior as well.

We have investigated the effect of detailed-balance violations on critical 
dynamics.
Generally, models with {\em non-conserved} order parameter are quite robust 
against non-equilibrium perturbations.
The relaxational models A and C only have an equilibrium fixed point.
For the SSS model, comprising models E and G, genuine non-equilibrium fixed
points do exist, corresponding to unidirectional couplings between order
parameter and conserved fields, but are unstable.
Thus, at criticality, the standard critical behavior is eventually recovered.
This remains true even when the conserved noise is rendered spatially 
anisotropic.
Essentially the same statements apply for models B, D, J and H with 
{\em conserved} order parameter, provided detailed-balance violations are 
introduced {\em isotropically}.
With spatially {\em anisotropic} order parameter noise correlations, however,
we find (to one-loop order) no RG fixed points for models J and H with 
reversible mode-coupling terms.
In contrast, the two-temperature relaxational models B and D are asymptotically
described by an effective equilibrium model, with characteristic anisotropic,
long-range correlations.

U.C.T. is grateful to Z. R\'acz for introducing him to this field, and
acknowledges support from the NSF (DMR-0075725) and the Jeffress Memorial Trust
(J-594). 
J.E.S. acknowledges support from the DFG (SFB 413/TP C6), and the EU 
(Marie Curie fellowship ERB-FMBI-CT 97-2816).
We thank E. Frey, H.K. Janssen, B. Schmittmann, F. Schwabl, and R.K.P. Zia for
illuminating discussions.


\begin{thebibliography}{99}
\bibitem{bearev} B.~Schmittmann and R.K.P.~Zia, 
                in {\em Phase Transitions and Critical Phenomena}, 
                eds. C.~Domb and J.L.~Lebowitz, Vol.~17
                (Academic Press, London, 1995).

\bibitem{intrev} T.~Halpin-Healy and Y.-C.~Zhang, 
		Phys. Rep. {\bf 254}, 215 (1995);
		J.~Krug, Adv. Phys. {\bf 46}, 139 (1997).

\bibitem{deprev} D.S.~Fisher, Phys. Rep. {\bf 301}, 113 (1998).

\bibitem{dirper} J. Marro and R. Dickman, 
		{\em Nonequilibrium phase transitions in lattice models} 
		(Cambridge University Press, Cambridge, 1999). 

\bibitem{hohhal} P.C.~Hohenberg and B.I.~Halperin, 
                Rev. Mod. Phys. {\bf 49}, 435 (1977).

\bibitem{stafth} See, e.g.,
                J.~Zinn-Justin, 
                {\em Quantum Field Theory and Critical Phenomena}
                (Clarendon Press, Oxford, 1989).

\bibitem{intcon} R.~Graham, {\em Springer Tracts in Modern Physics}, Vol. 66 
                (Springer-Verlag, Berlin, 1973);
		U.~Deker and F.~Haake, Phys. Rev. A {\bf 11}, 2043 (1975).

\bibitem{janded} C.~De~Dominicis, 
		J. Phys. (Paris) Colloq. C {\bf 1}, 247 (1976);
                H.K.~Janssen, 
                Z. Phys. B {\bf 23}, 377 (1976);
		R.~Bausch, H.K.~Janssen, and H.~Wagner,
                {\em ibid.} {\bf 24}, 113 (1976).

\bibitem{cubani} I.J.~Ketley and D.J.~Wallace, 
                J. Phys. A {\bf 6}, 1667 (1974);
                E.~Br\'ezin, J.C.~Le~Guillou, and J.~Zinn-Justin,
                Phys. Rev. B {\bf 10}, 892 (1974);
                T.~Nattermann and S.~Trimper,
                J. Phys. A {\bf 8}, 2000 (1975).

\bibitem{modsab} B.I.~Halperin, P.C.~Hohenberg, and S.-k.~Ma,
		Phys. Rev. Lett. {\bf 29}, 1548 (1972);
		Phys. Rev. B {\bf 10}, 139 (1974);
		C.~De~Dominicis, E.~Br\'ezin, and J.~Zinn-Justin,
		{\em ibid.} {\bf 12}, 4945 (1975).

\bibitem{uwezol} U.C.~T\"auber and Z.~R\'acz,
                Phys. Rev. E {\bf 55}, 4120 (1997);
		U.C.~T\"auber, J.E.~Santos, and Z.~R\'acz,
		Eur. Phys. J. B {\bf 7}, 309 (1999); 
		e: {\em ibid.} {\bf 9}, 56 (1999).

\bibitem{modelj} S.-k.~Ma and G.F.~Mazenko,
		Phys. Rev. Lett. {\bf 33}, 1383 (1974);
		Phys. Rev. B {\bf 11}, 4077 (1975).

\bibitem{fhaake} F.~Haake, M.~Lewenstein, and M.~Wilkens,
                Z. Phys. B {\bf 55}, 211 (1984).

\bibitem{grins1} G.~Grinstein, C.~Jayaprakash, and Y.~He,
                Phys. Rev. Lett. {\bf 55}, 2527 (1985).

\bibitem{kevbea} K.E.~Bassler and B.~Schmittmann,
                Phys. Rev. Lett. {\bf 73}, 3343 (1994).

\bibitem{bearoy} B.~Schmittmann and R.K.P.~Zia, 
                Phys. Rev. Lett. {\bf 66}, 357 (1991).

\bibitem{kevzol} K.E.~Bassler and Z.~R\'acz, 
                Phys. Rev. Lett. {\bf 73}, 1320 (1994); 
                Phys. Rev. E {\bf 52}, R9 (1995).

\bibitem{beate2} B.~Schmittmann, Europhys. Lett. {\bf 24}, 109 (1993).

\bibitem{uwevam} V.K.~Akkineni and U.C.~T\"auber (unpublished, 2001). 

\bibitem{modscd} E.~Br\'ezin and C.~De~Dominicis,
		Phys. Rev. B {\bf 12}, 4954 (1975);
		B.I.~Halperin, P.C.~Hohenberg, and S.-k.~Ma,
		{\em ibid.} {\bf 13}, 4119 (1976).
		
\bibitem{grins2} G.~Grinstein, C.~Jayaprakash, and J.E.S.~Socolar,
                Phys. Rev. E {\bf 48}, R643 (1993).

\bibitem{modseg} B.I.~Halperin, P.C.~Hohenberg, and E.D.~Siggia,
		Phys. Rev. B {\bf 13}, 1299 (1976);
		R.~Freedman and G.F.~Mazenko,
		{\em ibid.}, 4967.

\bibitem{sssmod} L.~Sasv\'ari. F.~Schwabl, and P.~Sz\'epfalusy,
		Physica {\bf 81 A}, 108 (1975);
		H.K.~Janssen, Z. Phys. B {\bf 26}, 187 (1977).

\bibitem{modssh} C.~De~Dominicis and L.~Peliti, 
		Phys. Rev. B {\bf 18}, 353 (1978).

\bibitem{modelh} %K.~Kawasaki, Ann. Phys. {\bf 61}, 1 (1970);
		E.D.~Siggia, B.I.~Halperin, and P.C.~Hohenberg,
		Phys. Rev. B {\bf 13}, 2110 (1976).

\bibitem{uwejai} U.C.~T\"auber and J.E.~Santos (unpublished, 2001).

\bibitem{dasrao} J.~Das, M.~Rao, and S.~Ramaswamy,
		e-print cond-mat/0104566 (2000).

\end{thebibliography}
\end{document}